\documentclass[pdftex,12pt]{article}
\usepackage{jheppub,amsmath,amsfonts,amsbsy,latexsym,jheppub,amssymb,amscd,amstext}
\usepackage{pst-node}

\pdfoutput=1

\def\be{\begin{equation}}
\def\ee{\end{equation}}
\def\bea#1\eea{\begin{align}#1\end{align}}
\def\pd{\partial}
\def\a{\alpha}
\def\b{\beta}
\def\g{\gamma}
\def\d{\delta}
\def\m{\mu}
\def\n{\nu}

\def\l{\lambda}

\def\r{\rho}

\def\s{\sigma}

\def\bma{\begin{pmatrix}}
\def\ema{\end{pmatrix}}

\def\bi{\begin{itemize}}
\def\ei{\end{itemize}}

\title{\boldmath Ghost-free higher derivative unimodular gravity.}

\author[a,b]{Enrique Alvarez}\note[b]{ Corresponding author.}
\author[a]{ and Sergio Gonzalez-Martin}

\affiliation[a]{Departamento de F\'{\i}sica Te\'orica and Instituto de F\'{\i}sica Te\'orica, IFT-UAM/CSIC\\Universidad Aut\'onoma, 20849 Madrid, Spain}

\emailAdd{enrique.alvarez@uam.es} 
\emailAdd{sergio.gonzalezm@uam.es}

\abstract{The unimodular version of the ghost-free higher derivative gravity is obtained. It is the unimodular reduction of some particular lagrangians quadratic in curvature.
}
\begin{document}

\maketitle
\flushbottom
\newpage

\section{Introduction}

Unimodular gravity is an interesting truncation of General Relativity , where the spacetime metric is restricted to be unimodular
\be
g\equiv
\text{det}~g_{\m\n}=-1
\ee
It is convenient to implement the truncation through the (non invertible) map
\be
g_{\m\n}\longrightarrow |g|^{-1/n}~g_{\m\n}
\ee
The resulting theory is not Diff invariant anymore, but only TDiff invariant. Transverse diffeomorphisms are those such that their generator is transverse, that is
\be
\pd_\m \xi^\m=0
\ee

The ensuing action of Unimodular Gravity (cf. \cite{AlvarezGMHVM} for a recent review with references to previous literature),  reads

\be
S_{UG}\equiv \int d^n x~{\cal L}_{UG}\equiv\
-M_P^{n-2}\int |g|^{1/n}\left(R+{(n-1)(n-2)\over 4 n^2}{g^{\m\n}\nabla_\m g\nabla_\n g\over g^2}\right)
\ee
It can be easily shown using Bianchi identities that the classical equations of motion (EM) of Unimodular Gravity coincide with those of General Relativity with an arbitrary cosmological constant. The main difference at this level between both theories is that a constant value for the matter potential energy does not weight at all, which solves part of the cosmological constant problem (namely why the cosmological constant is not much bigger that observed). This property is preserved by quantum corrections.\par

While the nature of the cosmological constant makes Unimodular Gravity an appealing alternative to General Relativity, it is still an effective field theory for low energies as it has the same problems with renormalizability.\par

On the other hand, it has been known since long \cite{Stelle} that quadratic theories of gravity are quite interesting. They are renormalizable (even asymptotically free) and they are in many senses the closest analogues to Yang-Mills theories. The problem however is with unitarity or, equivalently, with a ghostly state in the spectrum. This problem can in turn be traced to the quartic propagators, which contradict the K\"allen-Lehmann spectral representation.
\par
It has been suggested by Siegel \cite{Siegel}, however,  that string theory provides a natural way out, namely exponential falling off of the propagators, of the type
\be
L=\phi \Box \,e^{-{\frac{\Box}{2 M^2}}\,\phi+\phi T }
\ee
Building on these ideas, in \cite{Modesto,Biswas} general actions of the type
\be
S=\int d(vol)\,R_{\m\n \r\s} \,{\cal O}^{\m\n \r\s}_{\a\b\g\d}\,R^{\a\b\g\d}
\ee
(where ${\cal O}$ is a differential operator) in the quadratic approximation in the weak field expansion $		g_{\m\n}\equiv \eta_{\m\n}+\kappa\,h_{\m\n}$
	are analyzed.
Using the same notation as in  \cite{Biswas} this yields the action,
\bea
S=&-\int d^n x\bigg\{\dfrac{1}{2}\,h_{\m\n} a(\Box)\Box h^{\m\n}+h^\s_\m b(\Box)\pd_\s\pd_\n h^{\m\n}+ h\, c(\Box) \pd_\m\pd_\n h^{\m\n}+\nonumber\\
&+\dfrac{1}{2}\, h\, d(\Box) \Box h+ h^{\l\s} \dfrac{f(\Box)}{ \Box}\,\pd_\s\pd_\l\pd_\m\pd_\n h^{\m\n}\bigg\}\label{Biswas}
\eea
Here all coefficients $a,b,c,d,f$ are dimensionless functions of the d'Alembert operator.
\par
The aim of the present work is to see if it is possible to extend this idea to the unimodular theory. As is well-known \cite{ABGV} flat space unimodular gravity (UG) is a consistent theory propagating spin 2 only, with no admixture of spin zero, in the same foot as (and inequivalent to) Fierz-Pauli, with the curious property that it does not admit massive deformations.
\par
 We shall see in the sequel that all ghost-free quadratic theories admit unimodular cousins.
\section{Unimodular reduction}
In \cite{Alvarez} we have demonstrated that there is a (non injective) map which we call unimodular reduction, UR that yields the flat space unimodular theory out of the ordinary Fierz-Pauli one. Similar mappings can also be introduced also in the full non-linear case. To be specific, the unimodular reduction (UR)  
\be
UR:\quad\, h_{\m\n}\rightarrow h_{\m\n}-{1\over n}\,h\,\eta_{\m\n}
\ee
 of the action \eqref{Biswas} reads
\bea
&S_{UG}\equiv UR[S]=-\displaystyle\int d(vol)\bigg\{\dfrac{1}{2} h_{\m\n}\, a(\Box) \Box\, h^{\m\n}+h_\m^\s\, b(\Box)\pd_\s\pd_\n\, h^{\m\n}+h^{\l\s}\,{f(\Box)\over \Box}\,\pd_\s\pd_\l\pd_\m\pd_\n\,h^{\m\n}-\nonumber\\
&-{2\over n}\,h^{\a\b}\left( b(\Box)+f(\Box)\right)\pd_\a\pd_\b\,h+h\left[\dfrac{1}{ n^2} f(\Box)-\dfrac{1}{2n} a(\Box)+\dfrac{1}{n^2} b(\Box)\right]\Box\,h\bigg\}
\label{UGBiswas}\eea

Please note that the UG action is independent of $c(\Box)$ and $ d(\Box)$. Moreover, the local (two-derivatives)  unimodular gravity in \cite{ABGV} corresponds to constant values of the functions, namely
\bea
&a=\dfrac{1}{2}\nonumber\\
&b=-{1\over 2}\nonumber\\
&f=0
\eea

When computing the UR of the equations of motion (EM) it is important to realize \cite{ABGV} that the unimodular reduction does not commute with the variation; that is
\be
\left[UR,EM\right]\neq 0
\ee
From now on, we shall work in momentum space, where the different functions $a(\Box)$, etc in \eqref{UGBiswas} are functions of $k^2$.
Actually, the EM stemming from the action $S$ can be written as
\be
K_{\m\n\r\s} h^{\r\s}=0
\ee
where
\bea
&K_{\m\n\r\s}={a\over 4} k^2\left(\eta_{\m\r}\eta_{\n\s}+\eta_{\m\s}\eta_{\n\r}\right)+{b\over 4}\left(k_\s k_\n\eta_{\m\r}+k_\r k_\n\eta_{\m\s}+k_\m k_\s \eta_{\n\r}+k_\m k_\r\eta_{\n\s}\right)+\nonumber\\
&+{c\over 2}\left(\eta_{\r\s}k_\m k_\n + \eta_{\m\n} k_\r k_\s\right)+{d\over 2}k^2 \eta_{\m\n}\eta_{\r\s}+f{k_\m k_\n k_\r k_\s\over k^2}
\eea
It is important to note that the EM are symmetrized, id est,
\bea
&K_{\m\n\r\s}=K_{\r\s\m\n}\nonumber\\
&K_{\m\n\r\s}=K_{\n\m\r\s}
\eea
It is clear that the unimodular equations of motion cannot be the unimodular reduction of the Fierz-Pauli ones, since $c(\Box)$ does not disappear, whereas it is not even present in the UG action. To obtain the later, there is a general procedure explained in \cite{ABGV}. Define
\bea
&K_{\m\n}\equiv K_{\m\n\r\s}\, \eta^{\r\s}\nonumber\\
&K\equiv K_{\m\n}\,\eta^{\m\n}
\eea
Then
\bea
&K^{UG}_{\m\n\r\s}\equiv K_{\m\n\r\s}-{1\over n}\left(K_{\m\n} \eta_{\r\s}+K_{\r\s}\eta_{\m\n}-{1\over n}\,K\,\eta_{\m\n}\eta_{\r\s}\right)
\eea
Where this operator is built in such a way that it inherits the previous symmetries,
\bea
&K^{UG}_{\m\n\r\s}=K^{UG}_{\r\s\m\n}\nonumber\\
&K^{UG}_{\m\n\r\s}=K^{UG}_{\n\m\r\s}\eea

plus an extra one
\be
K^{UG}_{\m\n\r\s}\,\eta^{\r\s}=0
\ee
This yields
\bea
&K^{UG}_{\m\n\r\s}={1\over 4}a\, \left(\eta_{\m\r} \eta_{\n\s}+\eta_{\m\s}\eta_{\n\r}\right) k^2-{b+f\over n} \left(k_\m k_\n\, \eta_{\r\s}+k_\r k_\s\,\eta_{\m\n}\right)+\nonumber\\
&+{1\over 4} b\left(k_\r k_\n \eta_{\m\s}+k_\r k_\m\eta_{\n\s}+k_\s k_\n\eta_{\m\r}+k_\m k_\s \eta_{\n\r}\right)+\nonumber\\
&+ {2(b+f)-n a \over  2 n^2}\,k^2\, \eta_{\m\n}\,\eta_{\r\s}+ f\,{k_\m k_\n k_\r k_\s\over k^2}
\eea
It is plain that
\bea
&k^\m K_{\m\n\r\s}^{UG}= {a\over 4}\left(k_\r\eta_{\n\s}+k_\s\eta_{\m\r}\right) k^2-{b+f\over n}\left( k^2 k_\n \eta_{\r\s}+ k_\n k_\r k_\s\right)+\nonumber\\
&+{b\over 4}\left(2k_\s k_\r k_\n+ k^2 k_\r \eta_{\n\s}+ k^2 k_\s \eta_{\n\r}\right)+{2(b+f)-n a\over 2 n^2}\, k^2 k_\n \eta_{\r\s}+ f\, k_\r k_\n k_\s
\eea
The Bianchi identity implies that $K_{\m\n\r\s} $ is transverse, so that the source term in \cite{Biswas} must also be conserved 
\be
 \pd_\n T^{\m\n}=0
\ee
This means that the source term after UR, which is the traceless piece of the  energy-momentum tensor, namely
\be
T^{T}_{\m\n}\equiv T_{\m\n}-{1\over n} T \eta_{\m\n}
\ee
(where $T\equiv \eta^{\m\n} T_{\m\n}$) is not transverse anymore, but rather
\be
\pd_\n T^T_{\m\n}={1\over n}\pd_\m T
\ee

This is a nontrivial constraint, which is true only when the functions $a$ and $b$ are such that
\bea
&a+ b=0\nonumber
\eea
in which case the trace is given by
\be
T= \left( {(n-2)a+2(n-1) f\over 2 n} k^2  \eta_{\r\s}+{2(n-1) f -(n-2) a\over 2}\,k_\s k_\r \right) h^{\r\s}
\ee
There are no constraints in the unimodular case on  the function $f(k^2)$.

\section{Propagators}
\be
K^{UG}= a_1 P_1+ a_2 P_2 + a_s P_0^s+ a_w P_0^w+a_\times P_0^\times
\ee
where 
\bea
&a_1=0\nonumber\\
&a_2=\dfrac{1}{2}a\,k^2\nonumber\\
&a_s={2(n-1) f -(n-2) a\over 2 n^2}\,k^2=L k^2\nonumber\\
&a_w={2 (n-1)^2 f -(n-1)(n-2) a \over 2 n^2}\,\,k^2=(n-1)Lk^2=(n-1) a_s\nonumber\\
&a_\times=\sqrt{n-1}{(n-2)a-2 (n-1) f\over 2 n^2}\,k^2=-\sqrt{n-1}L k^2=-\sqrt{n-1}\, a_s
\eea
where we have defined
\be
L\equiv {2(n-1) f-(n-2) a\over 2 n^2}
\ee
The discriminant vanishes:
\be
\Delta\equiv a_s a_w-a_\times^2= 0
\ee
This means that we have to introduce a \emph{TDiff} gauge fixing
\be
K_1^\text{gf}\equiv \a_1 P_1
\ee
and besides another one for Weyl's symmetry, namely 
\be
K_2^\text{gf}\equiv \a_2 ( P_0^w+(n-1)P_0^s+\sqrt{n-1}P_0^\times)
\ee

\noindent The full operator is then
\be
K_{gf}^{UG}= \a_1 P_1+ a_2 P_2 + \left(a_s+ (n-1)\a_2\right) P_0^s+ \left(a_w+\a_2\right) P_0^w+\left(\a_2\sqrt{n-1}+a_\times\right) P_0^\times
\ee
\par

\noindent Using the formulas in the appendix, the propagator is given by
\bea
K^{-1}_{UG}=&{1\over \a_1}P_1+\bigg\{{2\over a}P_2+\nonumber\\
&+\dfrac{1}{n^2\a_2 L}\left( (Lk^2+(n-1)\a_2)P^w_0+(L(n-1)k^2 +\a_2)P_0^s+\sqrt{n-1}(Lk^2-\a_2)P_\times\right)\bigg\}{1\over k^2}
\eea

The  interaction energy between external, gauge invariant sources is a physical quantity. In our case is given by the coupling of the graviton to the traceless piece of the energy-momentum tensor
\be
\int d(vol)\,T^T_{\m\n} h^{\m\n}.
\ee
The source then is not transverse, but  rather 
\be
\pd_\m \left(T^T\right)^{\m\n}={1\over n}\pd^\n T.
\ee

The only projectors that do not vanish when sandwiched between physical sources are
\bea
&T^{T\m\n}\cdot\left(P_2\right)_{\m\n\r\s}\cdot T^{T\r\s}= T_{\m\n}^2-{1\over n-1}\,T^2
\\
&T^{T\m\n}\cdot\left(P_1\right)_{\m\n\r\s}\cdot T^{T\r\s}=0\\
&T^{T\m\n}\cdot\left(P_0^s\right)_{\m\n\r\s}\cdot T^{T\r\s}= \dfrac{1}{n^2(n-1)}T^2
\\
&T^{T\m\n}\cdot\left(P_0^x\right)_{\m\n\r\s}\cdot T^{T\r\s}=-\dfrac{2}{n^2\sqrt{n-1}}T^2
\\
&T^{T\m\n}\cdot\left(P_0^w\right)_{\m\n\r\s}\cdot T^{T\r\s}=\dfrac{1}{n^2}T^2
\eea

This yields the value for the free energy in the linear limit of our theory to be

\be
W=\dfrac{1}{k^2}\left(  {2\over a}\,T_{\m\n}^2 +\dfrac{a-2n^2L}{a(n-1)n^2L}\,T^2\right)
\ee
In order not to disagree with the classical solar system GR tests, this must be proportional to the well-known result (which is the Fierz-Pauli result, reproduced also by UG)
\be
W_{FP}=\dfrac{1}{k^2}\left( T_{\m\n}^2-\dfrac{1}{n-2}T^2\right)
\ee
This  determines
\be
L=-\dfrac{a(n-2)}{2n^2}
\ee
that is, $f$ is given in terms of $a$ through
\be
f(k)=0
	\ee

It is known that the function $f(z)$ must be an entire function of the complex variable $z$. This condition is trivially satisfied here.

\section{Conclusions.}

We have demonstrated here that the complete set of higher-derivative ghost-free theories have a related unimodular theory which can be easily obtained from the parent theory through unimodular reduction.

These theories, as also happens with the usual Unimodular Gravity constructed out of the Einstein-Hilbert lagrangian, do not couple the constant vacuum energy to gravity; which makes them an interesting alternative to the standard ones.
\par
 Further work is needed to investigate if one can get different physical predictions from the unimodular reduction than those of the parent theory. In spite of some work  \cite{Carmelo}, no differences were found at tree level between Unimodular Gravity and General Relativity other than the r\^ole of the vacuum energy just mentioned.
\par
Let us finish with a word of caution. The suggestion has been made  \cite{Shapiro} that in spite of the original claims there are hidden ghosts  in the quantum version of these theories. This question asks for a more detailed analysis. Let us point out, finally, that it is interesting to generalize the bootstrap mechanism to the unimodular case, in particular for higher order in curvature theories.
Some suggestions have been made in \cite{Deser}. Work on this topic will be reported in due time.

\section*{Acknowledgments}

 This work has been partially supported by the European Union FP7 ITN INVISIBLES (Marie Curie Actions, PITN- GA-2011- 289442)and (HPRN-CT-200-00148); COST action MP1405 (Quantum Structure of Spacetime), COST action MP1210 (The String Theory Universe) as well as by FPA2016-78645-P (MICINN, Spain), and S2009ESP-1473 (CA Madrid). This work is supported by the Spanish Research Agency (Agencia Estatal de Investigaci—n) through the grant IFT Centro de Excelencia Severo Ochoa SEV-2016-0597.

\newpage
\appendix
\section{Barnes' projectors}
 We start with the longitudinal and transverse projectors
\bea
&\theta_{\a\b}\equiv\eta_{\a\b}-{k_\a k_\b\over k^2}\nonumber\\ 
&\omega_{\a\b}\equiv {k_\a k_\b\over k^2}
\eea
They obey
\bea
&\theta_\m^\n+\omega_\m^\n=\d_\m^\n\nonumber\\
& \theta_\a^\b\theta_\b^\g=\theta_\a^\g\nonumber\\
&\omega_\a^\b \omega_\b^\g=\omega_\a^\g
\eea
as well as
\bea
& tr~\theta^\n_\m=n-1\nonumber\\
&tr~\omega^\n_\m=1
\eea
The four-indices projectors are
\bea
&(P_2)_{\m\n\r\s})\equiv {1\over 2}\left(\theta_{\m\r}\theta_{\n\s}+\theta_{\m\s}\theta_{\n\r}\right)-{1\over n-1}\theta_{\m\n} \theta_{\r\s}\nonumber\\
&(P_1)_{\m\n\r\s}\equiv{1\over 2}\left(\theta_{\m\r}\omega_{\n\s}+\theta_{\m\s}\omega_{\n\r}+\theta_{\n\r}\omega_{\m\s}+\theta_{\n\s}\omega_{\m\r}\right)\nonumber\\
&(P_0^s)_{\m\n\r\s}\equiv {1\over n-1}\theta_{\m\n}\theta_{\r\s}\nonumber\\
&(P_0^w)_{\m\n\r\s}\equiv \omega_{\m\n}\omega_{\r\s}\nonumber\\
&(P_0^{sw})_{\m\n\r\s}\equiv{1\over \sqrt{n-1}}\theta_{\m\n}\omega_{\r\s}\nonumber\\
&(P_0^{ws})_{\m\n\r\s}\equiv{1\over \sqrt{n-1}}\omega_{\m\n}\theta_{\r\s}
\eea
They obey
\bea
&P_i^a P_j^b=\d_{ij}\d^{ab} P_i^b\nonumber\\
&P_i^a P_j^{bc}=\d_{ij}\d^{ab}P_j^{ac}\nonumber\\
&P_i^{ab} P_j^c=\d_{ij}\d^{bc} P_j^{ac}\nonumber\\
&P_i^{ab} P_j^{cd}=\d_{ij}\d^{bc}\d^{ad} P_j^a
\eea
as well as
\bea
&tr~((P_2)_{\m\n\r\s}))\equiv \eta^{\m\n} (P_2)_{\m\n\r\s}=0\nonumber\\
&tr~((P_0^s)_{\m\n\r\s})\equiv \eta^{\m\n} (P_0^s)_{\m\n\r\s}=\theta_{\r\s}\nonumber\\
&tr~((P_0^w)_{\m\n\r\s})\equiv \eta^{\m\n} (P_0^w)_{\m\n\r\s}=\omega_{\r\s}\nonumber\\
&tr~((P_1)_{\m\n\r\s})\equiv \eta^{\m\n} (P_1)_{\m\n\r\s}=0\nonumber\\
&tr~((P_0^{sw})_{\m\n\r\s})\equiv \eta^{\m\n} (P_0^{sw})_{\m\n\r\s}=\sqrt{n-1}~\omega_{\r\s}\nonumber\\
&tr~((P_0^{ws})_{\m\n\r\s})\equiv \eta^{\m\n} (P_0^{ws})_{\m\n\r\s}={1\over \sqrt{n-1}}~\theta_{\r\s}\nonumber\\
&(P_2)_{\m\n}^{\r\s}+(P_1)_{\m\n}^{\r\s}+(P_0^w)_{\m\n}^{\r\s}+(P_0^s)_{\m\n}^{\r\s}={1\over 2}\left(\d_\m^\n \d_\r^\s+\d_\m^\s \d_\r^\n\right)
\eea

Any symmetric operator can be written symbolically as
\be
K= a_2 P_2 + a_1 P_1 + a_w P_0^w + a_s P_0^s + a_\times P_0^\times
\ee
(where $P_0^\times\equiv P_0^{ws}+P_0^{sw}$).
Then
\be
K^{-1}={1\over a_2}P_2+{1\over a_1} P_1 +{a_s\over a_s a_w - a_\times^2}P_0^w+{a_w\over a_s a_w - a_\times^2}P_0^s-{a_\times\over a_s a_w - a_\times^2}P_0^\times
\ee

Sometimes the action of those projectors on tracefree tensors is  needed. Defining the trecefree projector
\be
\left(P_{tr}\right)_{\r\s}\,^{\l\d}\equiv {1\over 2}\left(\d_\r^\l \d_\s^\d+\d_\r^\d \d_\s^\l\right)-{1\over n}\eta_{\r\s}\eta^{\l\d}
\ee

It is a fact that
\bea
&P_2 P_{tr}=P_2\nonumber\\
&P_0^s P_{tr}=P_0^s-{n-1\over n}P_0^s-{\sqrt{n-1}\over n}P_0^{sw}\nonumber\\
&P_0^w P_{tr}=P_0^w-{\sqrt{n-1}\over n}P_0^{ws}-{1\over n}P_0^w\nonumber\\
&P_1 P_{tr}=P_1\nonumber\\
&P_0^{sw} P_{tr}=P_0^{sw}-{\sqrt{n-1}\over n}P_0^{ws}-{1\over n}P_0^w\nonumber\\
&P_0^{ws} P_{tr}=P_0^{ws}-{\sqrt{n-1}\over n} P_0^{sw}-{n-1\over n} P_0^s
\eea
               
\end{document}